\documentclass[%
 reprint,
 superscriptaddress,
 amsmath,amssymb,
 aps,
 pra,
floatfix,
showkeys
]{revtex4-1}

\usepackage{graphicx}
\usepackage{dcolumn}
\usepackage{bm}
\usepackage{hyperref}
\hypersetup{colorlinks=true, citecolor=blue, urlcolor=blue, linkcolor=blue}

\begin{document}


\title{The Truncated Metallo-dielectric Omnidirectional Reflector: Collecting Single Photons in the Fundamental Gaussian Mode with 95\% Efficiency}

\author{Wancong Li}
\altaffiliation{These authors contributed equally to this work.}
\affiliation{School of Physics, Wuhan National Laboratory for Optoelectronics, Huazhong University of Science and Technology, Luoyu Road 1037, Wuhan, 430074, People's Republic of China}

\author{Luis Morales-Inostroza}
\altaffiliation{These authors contributed equally to this work.}
\affiliation{Max Planck Institute for the Science of Light, Staudtstra{\ss}e 2, 91058 Erlangen, Germany}

\author{Weiwang Xu}
\affiliation{School of Physics, Wuhan National Laboratory for Optoelectronics, Huazhong University of Science and Technology, Luoyu Road 1037, Wuhan, 430074, People's Republic of China}

\author{Pu Zhang}
\affiliation{School of Physics, Wuhan National Laboratory for Optoelectronics, Huazhong University of Science and Technology, Luoyu Road 1037, Wuhan, 430074, People's Republic of China}

\author{Jan Renger}
\affiliation{Max Planck Institute for the Science of Light, Staudtstra{\ss}e 2, 91058 Erlangen, Germany}

\author{Stephan G{\"o}tzinger}
\email[Corresponding author, Email: ]{stephan.goetzinger@mpl.mpg.de}
\affiliation{Max Planck Institute for the Science of Light, Staudtstra{\ss}e 2, 91058 Erlangen, Germany}
\affiliation{University of Erlangen-N{\"u}rnberg, Staudtstra{\ss}e 2,91058 Erlangen, Germany}
\affiliation{Graduate School in Advanced Optical Technologies (SAOT), FAU, D-91052 Erlangen, Germany}

\author{Xue-Wen Chen}
\email[Corresponding author, Email: ]{xuewen\_chen@hust.edu.cn}
\affiliation{School of Physics, Wuhan National Laboratory for Optoelectronics, Huazhong University of Science and Technology, Luoyu Road 1037, Wuhan, 430074, People's Republic of China}


\date{\today}

\begin{abstract}
    We propose a novel antenna structure which funnels single photons from a single emitter with unprecedented efficiency into a low-divergence fundamental Gaussian mode. Our device relies on the concept of creating an omnidirectional photonic bandgap to inhibit unwanted large-angle emission and to enhance small-angle defect-guided-mode emission. The new photon collection strategy is intuitively illustrated, rigorously verified and optimized by implementing an efficient body-of-revolution finite-difference time-domain method for in-plane dipole emitters. We investigate a few antenna designs to cover various boundary conditions posed by fabrication processes or material restrictions and theoretically demonstrate that collection efficiencies into the fundamental Gaussian mode exceeding $95\%$ are achievable. Our antennas are broadband, insensitive to fabrication imperfections and compatible with a variety of solid-state emitters such as organic molecules, quantum dots and defect centers in diamond. Unidirectional and low-divergence Gaussian-mode emission from a single emitter may enable the realization of a variety of photonic quantum computer architectures as well as highly efficient light-matter interfaces.
\end{abstract}

\maketitle


\section{Introduction}

Single-photon sources have been identified as central building blocks for quantum technologies, like quantum cryptography, photonic quantum computation, non-classical spectroscopy, and metrology  \cite{SinglePhotonSources,LinearOpticalQQ,SinglePhotonSourcesIntro,Qradiometry}. In particular atom-like emitters in the solid state offer a promising route and have therefore been investigated heavily in the past \cite{SolidStateEmitterReview, PlanarDielectricAntenna}. The big advantage of a solid state approach is the ability to precisely control the environment of an emitter through micro- and nano-structuring and thus to engineer the emission properties of an emitter \cite{EfficientSourceofSP,HighEfficientQD,StronCouplingSurfPlas,UltraSpontaneous}. Solid-state single quantum emitters have become highly efficient and versatile sources of single photons, reaching a certain level of maturity \cite{PlanarDielectricAntenna,NearOptimalSPS,LiuSPS,TowardsOptimalSPS}. However, efficient photon collection alone is for many applications not sufficient. Single-photon based quantum technologies will ultimately require bright and quasi-deterministic emission of single photons into a well-defined spatial mode. For many applications the fundamental Gaussian mode is required. Photons in this particular mode can be efficiently coupled to a single-mode fiber for a convenient delivery. As a matter of fact, coupling of single photons into a single-mode fiber is often not driven by convenience but rather by necessity. Long distance quantum communication and entanglement generation between distant stationary qubits \cite{loophole13Km} are just two examples. Fiber coupling is also particularly important for the scalability of quantum gates based on photon-photon interactions \cite{legendPaper,YamamotoIndisti,PanBoson1,PanBoson2} where any non-ideal mode overlap heavily compromises the gate performance and thus ultimately prevents upscaling. Current thresholds for the required total efficiency (including photon detection) are as large as  50\% or even 66\% depending on the specific implementation of an all optical quantum computer \cite{HowGood50,HowGood66}.

The existing strategies to simultaneously achieve high photon extraction efficiency and Gaussian-mode emission are so far mostly based on the coupling of a single emitter to the fundamental mode of a nanowire or to a microcavity mode. In the first case the mode is evolved into a free-space Gaussian mode via tapering the wire to a nanotip or by expanding it to a trumpet-like microstructure \cite{HighEfficientQD,ControllingNanowire,NanowireSPGaussian}. The fabrication of both structures requires rather elaborated etching techniques. Approaches based on vertical microcavities exploit the Purcell effect to enhance the cavity mode emission, which then reaches the far field through the outcoupling mirror. Photon extraction efficiencies over $90\%$ are predicted theoretically \cite{Richard2019Gated,Bouwmeester2018PhhysRevApplFiberCoupled} and up to $79\%$ have been experimentally demonstrated in a narraow spectral range \cite{NearOptimalSPS,OnDemandQDPan,Senellart2013NatCommBright}. The idea of using a microcavity has also been expanded to in-plane structures, such as circular Bragg gratings or bull's eye structures \cite{TowardsOptimalSPS,StronglyEntangledPhotons,CircularBragGrating,CircularBragGrating2,BUllsNV,CircularGratingRapaport,SPEnanoPosition}. In this case the cavity enhances the coupling to the in-plane cavity modes. Photons are then efficiently scattered by these geometries out of plane for collection. Recent reports have shown remarkable extraction efficiencies of $95\%$ in theory and of $85\%$ \cite{StronglyEntangledPhotons} or about $80\%$ \cite{OnDemandSemiPan} in experiments,out of which $\sim65\%$ could be coupled into a single-mode fiber \cite{OnDemandSemiPan}. From all photon collection methodologies planar antenna structures have so far shown the highest extraction efficiencies, single-photon detection rates \cite{PlanarDielectricAntenna,99planarXuewen,liu2014} and unprecedented total efficiencies \cite{LiuSPS}. However, the modal structure of the emission has so far prevented an efficient coupling of the collected photons to a single-mode fiber. Here, we propose a new photon collection strategy based on a truncated metallo-dielectric omnidirectional reflector which results in near-unity collection efficiency and simultaneously a nearly ideal Gaussian mode profile in the far field. A body-of-revolution finite-difference time-domain method \cite{ComputationalElectro} (BOR-FDTD) for in-plane dipole emitters is developed to rigorously verify the strategy and to efficiently optimize the structural parameters of the device. We apply the concept to various structures that can be readily fabricated. Additionally, we show that the devices are broadband and robust against fabrication imperfections.

\section{Theory and Design}
\begin{figure*}[htbp]
\centering
\includegraphics[width=\linewidth]{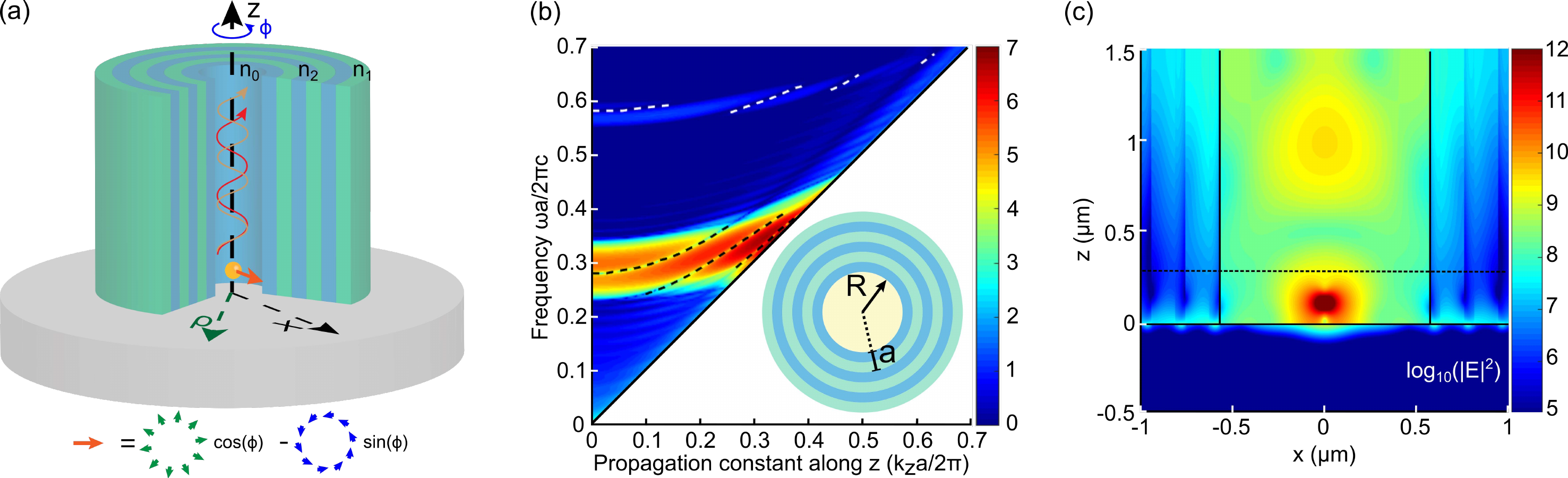}
\caption{(a) Schematic diagram of the truncated metallo-dielectric omnidirectional reflector, i.e., a hollow circular dielectric grating placed on a silver mirror. A quarter of the grating structure is removed to reveal the internal structure. An emitter positioned on the symmetry axis of the cylindrical coordinate system couples to the defect-guided modes. The illustration at the bottom shows that a vector along $\hat{x}$ on the symmetry axis can be expressed as a linear combination of $\hat{\rho}$ and $\hat{\phi}$. (b) Dispersion relation of an infinitely long hollow-core coaxial waveguide with radius $R = 3a$ ($a=t_1+t_2$, $t_1=130$ nm, $t_2=55$) nm, refractive indices  $2.58$/$1.38$ (blue/green layers in inset) and core refractive index $n_0=1.0$; blue regions indicate extended modes that propagate in the entire bilayer structure. The black-dashed (white-dashed) lines indicate the defect guided modes in the first (second) bandgap. See text for details. (c) Electric field intensity distribution in x-z plane obtained by the BOR-FDTD simulation with $m = 1$. An in-plane dipole is positioned in the semi-infinite coaxial waveguide at a distance of $120$ nm from the silver mirror. The black-dashed line indicates a possible height where the structure could be truncated to obtain a Gaussian emission profile.}
\label{fig:Fig1}
\end{figure*}
The truncated metallo-dielectric omnidirectional reflector is inspired by concepts known from photonic crystals \cite{InhibitesSE,StronDisorder}, specifically by strategies for inhibiting large-angle (in-plane) emission and exploiting photonic-crystal defect waveguides \cite{PhotonicCrystal} to enhance small-angle (out-of-plane) defect-mode emission. About two decades ago, Fink $et$ $al.$ proposed a type of one-dimensional periodic film stack capable of omnidirectional total reflection for all polarizations of incident light over a certain wavelength range\cite{OmnidirectionalTheo} and later demonstrated its application as a cladding of an all dielectric coaxial waveguide \cite{OmnidirectionalTheo2,OmnidirectionalTheo3}. Based on this idea, we propose a truncated metallo-dielectric omnidirectional reflector as illustrated in Figure \ref{fig:Fig1}(a). The enclosed central area with a radius of $R$ should have a low index of refraction denoted as $n_0$ (ideally $n_0 = 1$) and the dielectric bilayers have thicknesses of $t_1$ and $t_2$ and refractive of indices of $n_1$ and $n_2$, respectively. The circular periodic enclosing of the emitter and the low-index core force its emission into defect-guided modes. The stack of circular dielectric bilayers with a finite height $h$ is placed on a flat silver mirror. Photons can thus leave the structure only on one side. The truncation of the structure at an appropriate height $h$ enables efficient conversion of the defect-guided modes into the fundamental Gaussian mode in the upper homogeneous medium with a refractive index of $n_3$. We note that the working principle as described above completely differs from circular gratings which have been used previously for efficient photon collection. For example, our device does not rely on  Purcell enhancement of the spontaneous emission rate.

To set the ground for our analysis, we first assume the height of the circular bilayer stack to be infinite. This situation is equivalent to a dipole radiating inside an all dielectric coaxial waveguide \cite{OmnidirectionalTheo2,OmnidirectionalTheo3}. A necessary condition for omnidirectional reflection is that light in the core region with an index $n_0$ cannot access the Brewster angle of the bilayer \cite{OmnidirectionalTheo} $\theta_b = \tan^{-1}(n_2/n_1)$, which requires that $n_0<n_1 n_2/\sqrt{n_1^2+n_2^2}$ ($n_2>n_1$). Otherwise there would exist at least a $p$-polarized plane wave that refracts at the $n_0/n_2$ interface and then transmits through the entire ($n_2/n_1$) bilayer stack. In the case of a single emitter embedded in a high index material (like epitaxially grown InGaAs quantum dots) this means that one has to partially remove the material around the emitter in order to effectively create a low index $n_0$. To inhibit the coupling of the dipolar emission into unwanted modes having a large in-plane wavenumber, we design the bilayer in such a way that the Bragg reflection condition is satisfied for the modes with small longitudinal wavevectors $k_z$ (i.e., large in-plane wavenumbers). Specifically, each layer should become a quarter-wave layer so that the target wavelength will be in the bandgap \cite{PCModeling}, i.e.,
\begin{eqnarray}
t_1 \sqrt{n_1^2 - (k_z/k_0)^2}&=&\lambda_0/4,\\
t_2 \sqrt{n_2^2 - (k_z/k_0)^2}&=&\lambda_0/4,
\label{eq:eq1}
\end{eqnarray}
where $\lambda_0$ is the central vacuum wavelength of interest and $k_0$ is the wavenumber in vacuum. $t_i$ and $n_i$ ($i=1,2$) are the thickness and refractive index of material $i$ of the bilayer. For an emitter with in-plane dipole orientation, we find that $k_z= 0.5k_0$ to be a good choice. The central core radius $R$ determines the number of defect waveguide modes an emitter can couple to. For small values of $R$ a single-mode hollow omniguide is possible \cite{OmnidirectionalTheo2}. However, for our application it turns out that the mode size is too small and the beam thus diverges too strongly when exiting the structure. Therefore, we have considered a core size of $R \sim 3(t_1 +t_2)$ to support two to three defect-guided modes in the wavelength range of interest, which are then converted into a low-divergence fundamental Gaussian mode if the  grating structure is truncated at the correct height.

To study the mode properties of the all dielectric coaxial waveguide, we apply a transfer matrix method \cite{PCModeling,Pochi1978} with appropriate boundary conditions and solve the dispersion equation. As a concrete example, we consider a stack of 10 bilayers with a thickness of $t_1=130$ nm and $t_2=55$ nm for a target wavelength of $\lambda_0 = 640$ nm. The refractive indices are $n_0 = 1.0$, $n_1 = 1.38$ (e.g., MgF$_2$) and $n_2 = 2.58$ (e.g., TiO$_2$). Note that the innermost ring is necessarily made out of the material with a larger refractive index which is in our case $n_2$. Figure \ref{fig:Fig1}(b) displays a color-mapped graph for a determinant derived from the dispersion relation equation (Eq. (S13) in SM) as a function of the normalized frequency and the propagation constant in $z$ direction (see Supporting Information for further details). The determinant is plotted in a logarithmic scale and a small magnitude means that there exists an electromagnetic mode in the structure. Here we plot only the upper left side of the light line (thick black line), where the modes can propagate along the z direction in the low-index core. The continuous blue region indicates the modes that exist in the entire bilayer stack. The black-dashed (white-dashed) lines inside the red-yellowish (light-blue) background region are the defect waveguide modes in the first (second)  photonic bandgap. Single emitters embedded in the core region with an appropriate emission spectrum couple to these defect-guided modes.

The above analysis of the infinitely long all-dielectric coaxial waveguide only serves as a starting point for the design of the proposed truncated metallo-dielectric omnidirectional reflector, where the grating has only a sub-wavelength height and a metallic mirror underneath as a reflector. To quantitatively study a dipole radiating in such a complicated structure, one has to resort to three-dimensional (3D) electromagnetic numerical calculations, which are computationally intensive tasks since the simulation domain is about $\sim 1000\lambda_0^3$. It becomes even more demanding when there are quite a few geometric parameters to be optimized. For the purpose of demonstrating the principle and the device optimization, we restrict ourselves at first to an emitter located on the symmetry axis of the rotationally symmetric structure and later check the properties for an emitter placed off axis. In particular, we note that a linearly-polarized in-plane dipole, which we assume without loss of generality orients along the $\hat{x}$ direction, can be decomposed into a linear combination of dipoles along the unit vectors $\hat{\rho}$ and $\hat{\phi}$ of a polar coordinate system (see Figure \ref{fig:Fig1}(a)):
\begin{equation}
p_0 \hat{x} = p_0 \cos(\phi) \hat{\rho}  - p_0 \sin(\phi) \hat{\phi} .
\label{eq:eq2}
\end{equation}

Here $p_0$ is the magnitude of the dipole moment. With the chosen excitation source and a rotationally symmetric structure, all components of the electromagnetic field ($E_\rho$,\allowbreak$E_z$,\allowbreak$E_\phi$,\allowbreak$H_\rho$,\allowbreak$H_z$,\allowbreak$H_\phi$ ) have a dependence on $\phi$ in the form of $\psi_c(\rho,z)\cos(\phi) + \psi_s(\rho,z)\sin(\phi)$, which allows us to develop a BOR-FDTD method with azimuthal number $m = 1$ \cite{ComputationalElectro,BORtheo}. This method effectively reduces the computational demand of a 3D problem to that of a 2D case, consequently enabling us to perform rigorous numerical simulations and an optimization of the structural parameters in an efficient manner. The validity and excellent accuracy of the method have been benchmarked with a commercial 3D FDTD solver as shown in the Supporting Information. We have applied this BOR-FDTD method to compute the fields due to an in-plane dipole radiating inside a semi-infinite all-dielectric coaxial waveguide structure placed on a silver mirror \cite{AgDispersion}. The parameters of the all-dielectric coaxial waveguide structure are the same as in Figure \ref{fig:Fig1}(b). Figure \ref{fig:Fig1}(c) displays the total electric field distribution near the hollow core region for the in-plane dipole source which is placed at a distance of $120$ nm from the silver mirror. One observes that the radiation of the dipole is well confined in the core region and the field distribution evolves periodically along $z$ direction. The field distribution through the black-dashed line resembles a Gaussian profile, which motivates us to truncate the grating at an appropriate position $z = h$ to obtain a Gaussian-mode emission into free space.

To quantify the efficiency of the truncated omnidirectional reflector for directing single photons, we introduce two parameters, the collection efficiency $\gamma$ defined as the ratio of the far-field emission to the total emission from the emitter, and $\eta$ as the projection efficiency of the far-field emission onto the fundamental Gaussian mode. Thus, the total efficiency of the emission into the Gaussian mode is given by $\gamma \eta$. To evaluate $\eta$, we decompose the numerically-computed electric field $\Vec{E}(\rho,\phi,z)$ in medium $n_3$ into Laguerre-Gaussian modes which form a complete orthogonal basis \cite{Decomposition}. As we show in the Supporting Information, the normalized coefficient of the fundamental Gaussian mode can be expressed as 
\begin{equation}
c_1 = \frac{\iint^{}_\Omega ds \left[ \Vec{E}(\rho,\phi,z)\ast \Vec{E^{\ast}_g}(\rho,\phi,z-z_0,w_0) \right]}{\sqrt{\iint^{}_\Omega ds \left|\Vec{E}(\rho,\phi,z) \right|^2}\sqrt{\iint^{}_\Omega ds \left|\Vec{E_g}(\rho,\phi,z-z_0,w_0) \right|^2}},
\label{eq:eq3}
\end{equation}
Here $\Vec{E_g}(\rho,\phi,z-z_0,w_0)$ is the field distribution of the fundamental Gaussian mode where $(\rho,\phi,z)$ are cylindrical coordinates, $w_0$ is the beam waist and $z_0$ is the position of the beam waist in the cylindrical coordinate system at hand. The integration is performed over the whole plane at a fixed z. Note that the coefficient $c_1$ doesn’t depend on $z$ but is a function of $w_0$ and $z_0$ (see Supporting Information for further details). By scanning $w_0$ and $z_0$, we obtain a maximum value of $|c_1|$ for the most appropriate set of the Laguerre-Gaussian modes. The projection efficiency $\eta$ is given by $max \left\lbrace |c_1 |^2 \right\rbrace$. With the above numerical method and evaluation criteria, we can explore various designs to optimize the truncated omnidirectional reflector. In the following sections, we present five designs to illustrate the good performance and generality of the proposed device. The design procedure can be described as follows: 1) we first determine the thickness of the two bilayer materials by satisfying the quarter-wave condition. This bilayer thickness serves as an initial guess for the parameters since the presence of the metal layer and the top layer medium modifies the dispersion relation. 2) we optimize the thickness of each bilayer material, the truncation height $h$, the dipole position $d$ and the radius of the defect $R$ via efficient BOR-FDTD calculations with the goal of reaching the highest total efficiency $\gamma \eta$.

\section{Results and Discussion}
\begin{figure*}[htb]
\centering
\includegraphics[width=0.69\linewidth]{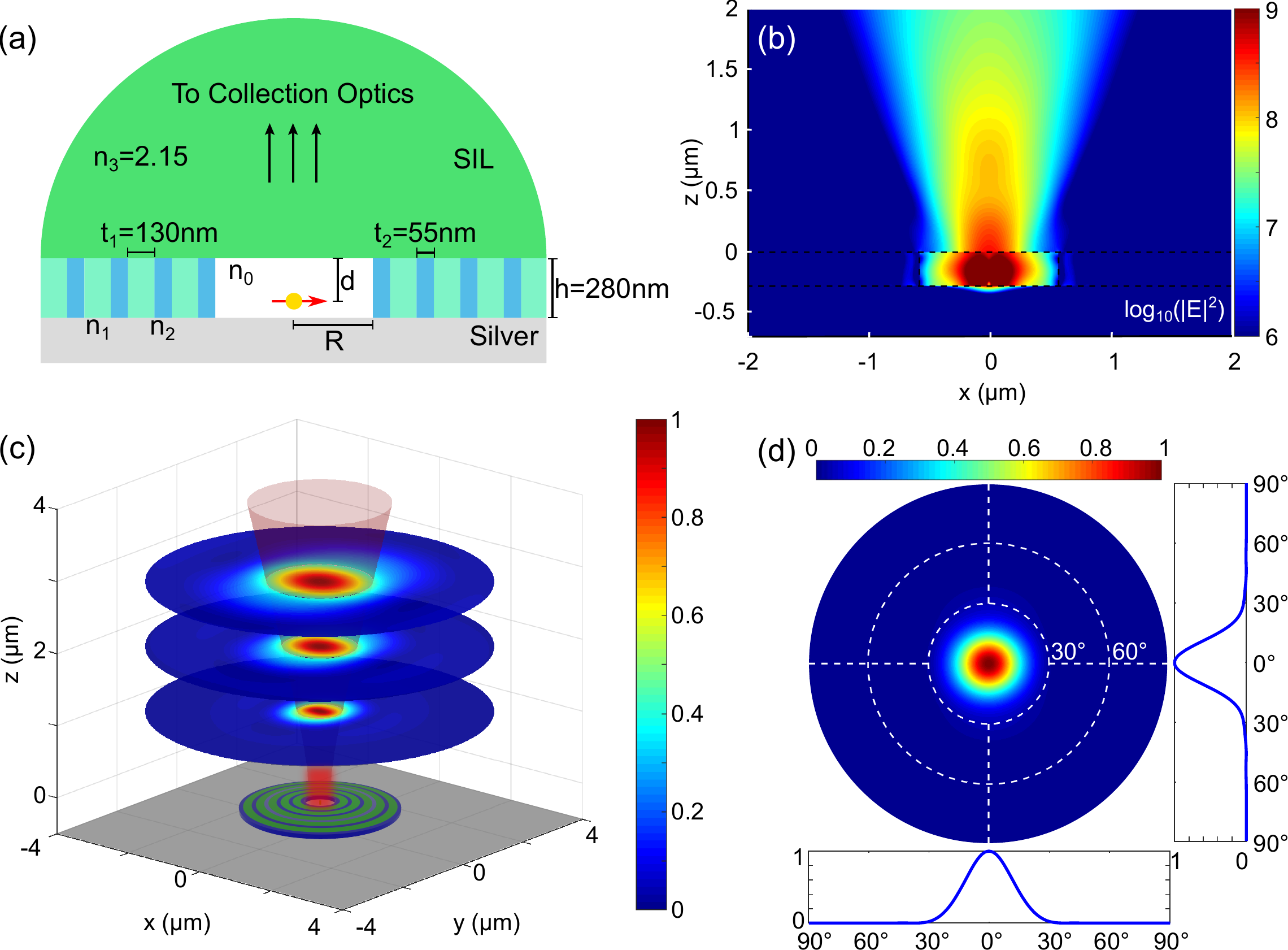}
\caption{An ideal truncated omnidirectional reflector for directing single photons. (a) Sketch of the device. The top medium has a refractive index of $n_3=2.15$. The optimized device parameters are $h = 280$ nm, $d = 160$ nm and $R = 570$ nm, respectively. The other parameters are the same as in Figure \ref{fig:Fig1}(b). (b) Electric field intensity distribution in a logarithmic scale for an in-plane dipole radiating inside the structure. (c). Intensity distributions in horizontal planes at three different heights. The fields resemble the free propagation of a Gaussian mode as indicated by the red beam. (d) Far-field emission pattern of the emitter as a function of in-plane wavenumbers.}
\label{fig:Fig2}
\end{figure*}

In the first and most basic example we place an in-plane dipole without any surrounding medium inside the hollow core region as schematically shown in Figure \ref{fig:Fig2}(a). The same parameters of the all dielectric waveguide structure as in Figure \ref{fig:Fig1}(b) are used. The structure is truncated at a height $h$ and the top layer medium has a refractive index of $n_3 = 2.15$ (e.g., ZrO$_2$). The top medium may have the form of a half sphere, which functions as a so-called solid immersion lens (SIL). It avoids refraction and thus lowers the divergence angle of photons leaving the device. The position of the dipole with respect to the SIL's flat surface is denoted as $d$. We obtain the optimal performance for $h = 280$ nm and $d = 160$ nm. Figure \ref{fig:Fig2}(b) displays the electric field intensity distribution (logarithmic scale) in the vertical plane for a dipole radiating in the structure. One sees that although there is some field penetrating vertically into the metal layer and transversely into the circular grating area the vast majority of the intensity is confined in the core region and unidirectionally propagates upwards for collection. Figure \ref{fig:Fig2}(c) depicts the intensity distribution in the horizontal plane at three different heights in medium $n_3$, which resembles the free propagation of a Gaussian beam. We have applied the Lorentz reciprocity theorem to compute the far-field emission pattern of the emitter in the planar multilayer system from the electric and magnetic fields calculated via the BOR-FDTD \cite{NearToFarfield,EmissionPattern}. The far-field emission pattern shows a well-confined single central lobe as plotted in Figure \ref{fig:Fig2}(d), which clearly resembles a Gaussian distribution in $k$-space. The divergence angle $\theta$, where the intensity falls to $1/e^2$, is $23.6^{\circ}$. Our calculations result in $\gamma=97.9\%$ and $\eta=97.3\%$, which amounts to a total efficiency of $95.1\%$ into the fundamental Gaussian mode. We also note that the Purcell factor for our device is very close to one.

\begin{figure*}[htbp]
  \centering
  \includegraphics[width=0.7\linewidth]{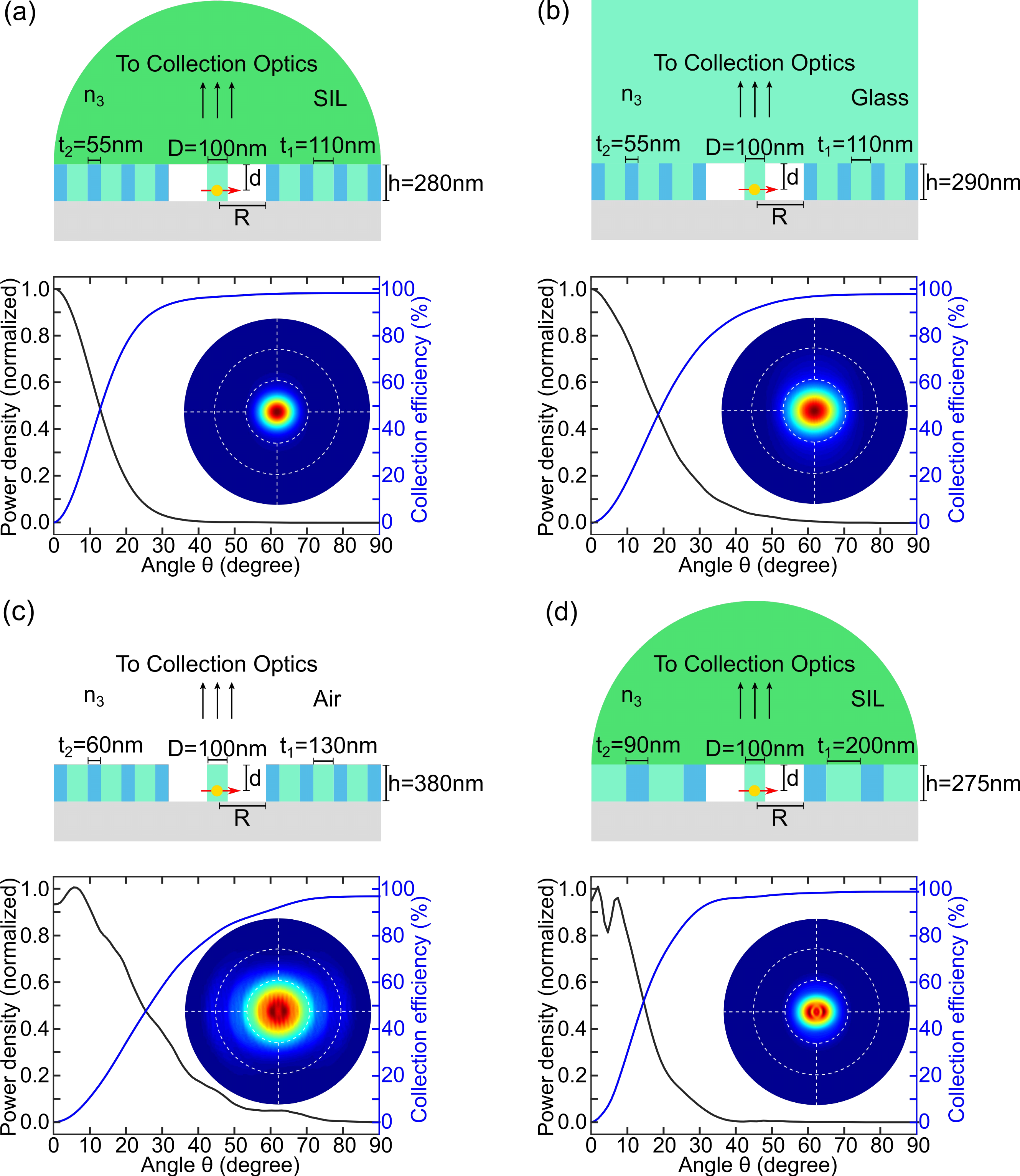}
  \caption{Optimized truncated omnidirectional reflectors for various experimental situations. The upper part of each figure is the sketch of the device structure and the lower part shows the angular emission density and angle-dependent collection efficiency while the inset is displaying the 2D angular emission pattern. In all cases, $n_2 = 2.58$ and $n_1 = n_e = 1.5$ with $n_e$ being the index of the $100$ nm-diameter nanodisk at the center of the device, where the emitter is embedded. The optimized values of the bilayer thicknesses and height are shown in the sketch above each graph. (a) $n_3 = 2.15$. Optimal structural parameters are $d = 145$ nm, and $R = 590$ nm. Performance: $\gamma=98.1\%$ and $\eta=96.9\%$. (b) $n_3 = 1.5$. Optimal structural parameters are $d = 180$ nm, and $R = 600$ nm. Performance: $\gamma=97.8\%$ and $\eta=90.3\%$. (c) $n_3 = 1.0$. Optimal structure parameters are $d = 265$ nm and $R = 500$ nm. Performance $\gamma=96.2\%$ and $\eta=90.7\%$. (d) Use of the second photonic bandgap shown in Figure \ref{fig:Fig1}  and $n_3 = 2.15$. The optimized structural parameters are $d = 140$ nm and $R = 550$ nm. Performance: $\gamma=97.9\%$ and $\eta=96.2\%$.}
  \label{fig:Fig3}
\end{figure*}

The second design considers a more realistic case when the emitter is embedded in a solid medium with a refractive index $n_e=1.5$ as indicated in the upper part of Figure \ref{fig:Fig3}(a). To satisfy the necessary condition of omnidirectional reflection $n_0<n_1 n_2/\sqrt{n_1^2+n_2^2}$, we remove part of the medium around the emitter to effectively decrease the refractive index $n_e$. As shown in the schematic diagram, the central core is almost empty, except for a nanodisk with a diameter of $100$ nm and thickness equal to the height of the grating structure. Here we chose to keep the refractive index $n_3 = 2.15$ and change the refractive index $n_1$ to $1.5$. Possible material choices of $n_1$ include silicon oxide or a polymer like polymethyl methacrylate (PMMA). This leads to a change in the thickness of the bilayers ($t_1= 110$ nm, $t_2=50$ nm) and the optimized parameters are found to be $h = 280$ nm, $d=145$ nm and $R = 590$ nm respectively. The color-coded traces in Figure \ref{fig:Fig3}(a) illustrate the angle-dependent emission power density and collection efficiency for the optimized device. The inset displays the angular emission pattern which has a single central lobe with a divergence angle of $24.5^{\circ}$. The calculations show $\gamma=98.1\%$ and $\eta=96.9\%$ for this case (total efficiency $\gamma \eta =95.1\%$). Note that a high refractive index SIL keeps the divergence angle low and therefore makes this structure for example compatible with a cryogenic environment. Figure \ref{fig:Fig3}(b) displays a structure where the top layer is a simple microscopy cover glass with $n_3 = 1.5$. As shown by the inset and lower panel of Figure \ref{fig:Fig3}(b), the emission remains basically Gaussian, however with a larger divergence angle of $37.3^{\circ}$. The calculations show $\gamma = 97.8\%$ and $\eta=90.3\%$ for this case. This omnidirectional antenna reflector is designed to be combined with an oil immersion collection optics, which can easily cover the increased divergence angle. One can also imagine scenarios where the application of a high-refractive index top layer is not possible or unwanted. Figure \ref{fig:Fig3}(c) shows the performance of such a structure with $n_3 = 1.0$. Both collection efficiency and projection efficiency are still remarkably high ($\gamma=96.2\%$, $\eta=90.7\%$). The emission profile gets slightly distorted and the divergence angle is with $50^{\circ}$ rather large. However, an air objective with a numerical aperture of $0.95$ would still be able to capture the the emitted photons. 

The three examples discussed above rely on the coupling to the defect modes in the first photonic bandgap of the all-dielectric coaxial waveguide, which provide good performance but at the same time require quite small feature sizes. However, one may also consider the use of defect modes in the second bandgap of the structure, as shown by the white-dashed lines in the upper left corner of Figure \ref{fig:Fig1}(b). The fourth example given here explores the above idea by expanding the size of the bilayer (e.g. PMMA/TiO$_2$) by a factor of $\sim 2$ (with some adjustments from optimization), i.e., $t_1 = 200$ nm, $t_2 = 90$ nm as shown in the upper panel of Figure \ref{fig:Fig3}(d). Remarkably, the calculations result in $\gamma = 97.9\%$ and $\eta=96.2\%$ with a small divergence angle of $26.4^{\circ}$. The emission pattern and angle-dependent collection efficiency are displayed in Figure \ref{fig:Fig3}(d). The advantage of this design is that the pitch sizes have doubled as compared to previous examples, easing the demand on nanofabrication.

\begin{figure}[htb]
  \centering
  \includegraphics[width=\linewidth]{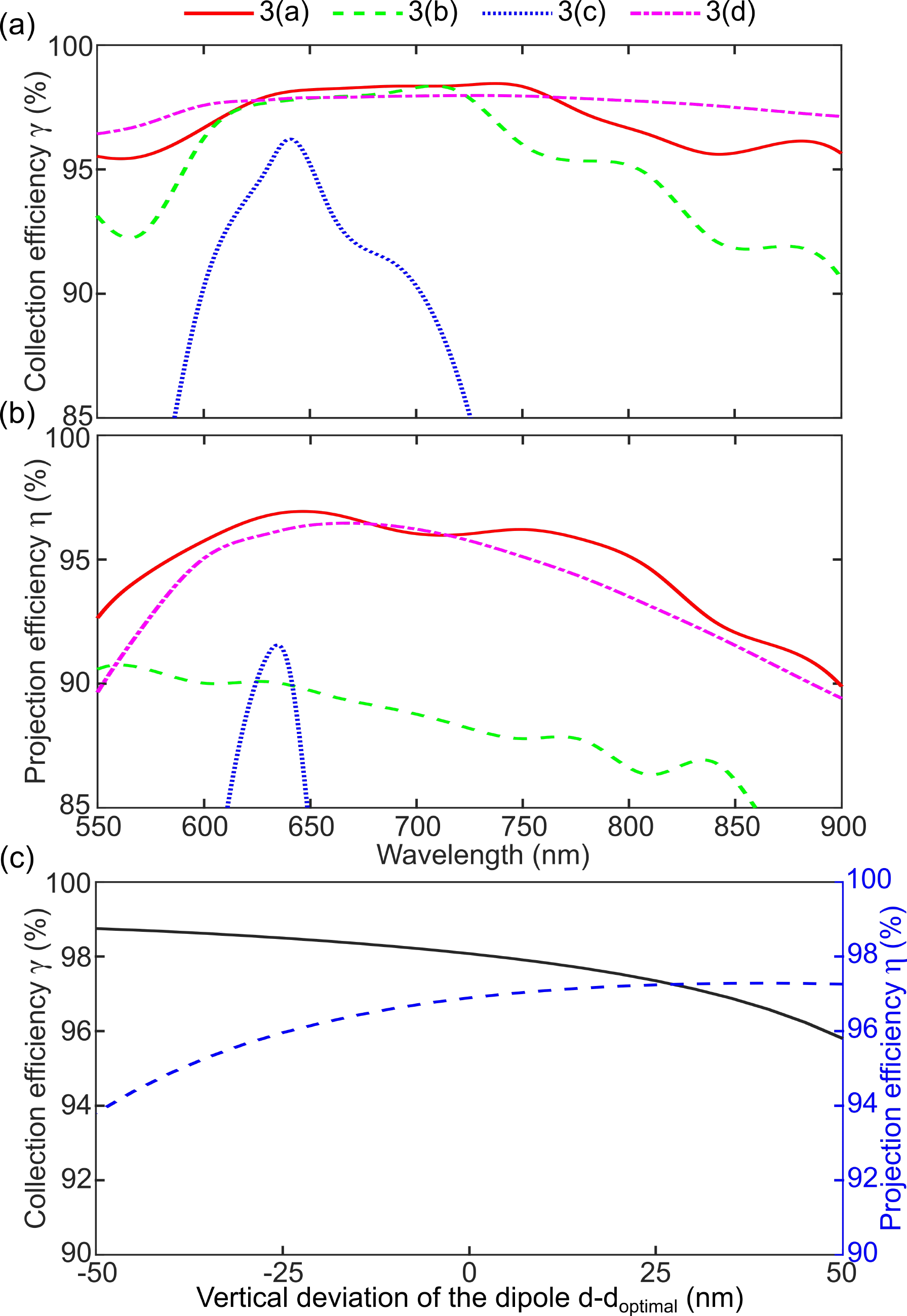}
  \caption{(a) Collection efficiency and (b) projection efficiency  into the fundamental Gaussian mode as a function of wavelength for the four structures studied in Figure \ref{fig:Fig3}. (c) Collection efficiency and projection efficiency as a function of the deviation from the optimal dipole position in the vertical direction for the device in Figure \ref{fig:Fig3}(a). The optimal position is the place where the total efficiency is maximum.
  }
  \label{fig:Fig4}
  \end{figure}

All four designs presented in Figure \ref{fig:Fig3} are optimized based on the concept of inhibiting the in-plane emission and enhancing guided-defect modes without the need of Purcell enhancement. Indeed, the calculated Purcell factors are $1$, $0.95$, $0.73$ and $1.01$ for cases studied in Figure \ref{fig:Fig3}(a), \ref{fig:Fig3}(b), \ref{fig:Fig3}(c), and \ref{fig:Fig3}(d), respectively. In the following, we examine the spectral properties and the tolerance to the emitter position.  Figure \ref{fig:Fig4}(a) and \ref{fig:Fig4}(b) summarize the spectral responses of the collection efficiencies and the projection efficiencies for the four device structures studied in Figure \ref{fig:Fig3}. All of the studied structures show more than $97\%$ collection efficiency over a broad spectral range except for the case with $n_3 =1.0$, which has a slightly lower collection efficiency. This range exceeds 200 nm for the two cases with $n_3 = 2.15$ and reduces to about $150$ nm for the case with $n_3 = 1.5$. For the case with $n_3 =1.0$, a collection efficiency over $90\%$ is achievable over $100$ nm bandwidth. All the studied structures exhibit pretty large projection efficiencies into the fundamental Gaussian mode over a broad spectral range as shown in Figure \ref{fig:Fig4}(b). Particularly for the two cases with $n_3 = 2.15$, the range with $\eta >95\%$ is more than $150$ nm. One may note that a higher refractive index of the top medium modifies the emission in a way that leads to both a larger collection efficiency and projection efficiency. The broadband responses of our devices may originate from the subwavelength truncation of the grating structure. Finally, we have studied exemplarily the dependence of the emission properties on the longitudinal positions of the dipole for the device shown in Figure \ref{fig:Fig3}(a). Figure \ref{fig:Fig4}(c) displays $\gamma$ and $\eta$ as a function of the dipole position $d$. One observes that within a deviation of $\pm 25$ nm from the optimal dipole position the collection efficiency and projection efficiency change by only $0.6\%$ and $0.7\%$, respectively. We also performed full 3D-FDTD simulations or a laterally displaced emitter. Our calculations show that deviations as large as 25 nm cause changes of $0.01\%$ and $0.4\%$ to the collection efficiency and projection efficiency, respectively. This demonstrates that the device performance is quite robust against deviations of the emitter from the optimal position and  thus relaxes the requirements on fabrication precision. In all of the devices presented, we have used silver as material of the metal mirror. Possible problems with oxidization can be minimized by depositing a few nanometers of Al$_2$O$_3$ via atomic layer deposition. However, we have achieved similar performances for all of the devices studied (collection efficiency about $1\sim2\%$ lower) by implementing a gold bottom reflector as shown in Supporting Information. Our strategy and optimization protocol are applicable to a variety of solid-state single quantum emitters and material systems. In Section 7 of Supporting Information, we have explicitly provided optimal antenna designs based on defect modes in the second bandgap of the circular dielectric grating for single molecules doped in host nanocrystals \cite{Pazzagli2018ACSNanoSelfAssembled,Hadi2019NatCommNanoprinting}, nanodiamond with single color centers in a polymer matrix and self-assembled InGaAs quantum dots in a high index membrane. Nanofabrication strategies for the antenna devices are also discussed with graphical illustrations in Section 7 of the Supporting Information. For self-assembled InGaAs quantum dots, nanofabrication of the host material may degrade the spectral properties of the quantum dots \cite{Kartik2018PhysRevApplSingleSelf}, which can be remedied by applying surface passivation approaches \cite{Kartik2018PhysRevApplSingleSelf} and by controlling the charge noise via electric gating  \cite{Richard2015NatCommTransform,Richard2019Gated}. The antenna design can be adjusted to take practical requirements into account, such as the use of conductive transparent materials or atomic-layer deposition of additional materials around the emitter.

\section{Conclusions}
\label{sec:conclusions}

In summary, we have proposed a novel truncated metallo-dielectric omnidirectional reflector structure for highly efficient extraction of single photons, which reside in the fundamental Gaussian mode with a total efficiency exceeding $95\%$. The antenna is compatible with various material systems and operation conditions. Our designs are blueprints for the realization of single-photon sources able to deliver more than a hundred million photons per second into a single-mode fiber. Such a device would enable the realization of a variety of quantum computer architectures, like boson samplers \cite{BosonSamplePan} or cluster-state quantum computers \cite{OneWayQC,Schwartz434} with an unprecedented number of photons available. One can even speculate that the combination of several of these sources could demonstrate quantum supremacy with photons. The success of the later will not depend on the photon collection methodology but rather on the properties of the emitter itself. Furthermore, the unidirectional Gaussian-mode emission of the emitter opens the door for perfect spatial mode-matching between an illumination source and an emitter, leading to highly efficient light-matter interfaces \cite{PerfectReflectionDipole,SinglePhotonAbsorption,FullInversionTLS}.

\begin{acknowledgments}
    We acknowledge financial support from the National Natural Science Foundation of China (grant 11874166, 11604109);Thousand-Talent (Young) Program of China.
    This work was supported by the Max Planck Society. We thank Vahid Sandoghdar for fruitful discussions and continuous support.
\end{acknowledgments}

\section*{Notes}
The authors declare the following competing financial interest(s): The approach presented in this paper is the subject of a patent application.



\bibliographystyle{apsrev4-1}
\bibliography{OMDA_arXiv_V1}

\end{document}